# The Abundance of Boron in Evolved A- and B-Type Stars

Kim A. Venn[1,2], David L. Lambert[3], and Michael Lemke[4]

[1] Univ. Sternwarte-München, Scheinerstr. 1, 81679 München, Germany
[2] Max-Planck-Institut für Astrophysik, 85740 Garching, Germany
[3] Univ. Texas at Austin, Dept. of Astronomy, RLM 15.308, Austin, TX, 78712, USA
[4] Institute of Astronomy, Madingley Road, Cambridge, CB3 0HA, England

**Abstract.** Boron abundances in A- and B-type stars may be a successful way to track evolutionary effects in these hot stars. The light elements – Li, Be, and B – are tracers of exposure to temperatures more moderate than those in which the H-burning CN-cycle operates. Thus, any exposure of surface stellar layers to deeper layers will affect these light element abundances. Li and Be are used in this role in investigations of evolutionary processes in cool stars, but are not observable in hotter stars. An investigation of boron, however, is possible through the B II 1362 Å resonance line.

We have gathered high resolution spectra from the IUE database of A- and B-type stars near 10 M$_\odot$ for which nitrogen abundances have been determined (by Gies & Lambert 1992 and Venn 1995). The B II 1362 Å line is blended throughout the temperature range of this program, requiring spectrum syntheses to recover the boron abundances. For no star could we synthesize the 1362 Å region using the meteoritic/solar boron abundance of log$\epsilon$(B)=2.88 (Anders & Grevesse 1989); a lower boron abundance was necessary which may reflect evolutionary effects (e.g., mass loss or mixing near the main-sequence), the initial composition of the star forming regions, or a systematic error in the analyses (e.g., non-LTE effects). Regardless of the initial boron abundance, and despite the possibility of non-LTE effects, it seems clear that boron is severely depleted in some stars. It may be that the nitrogen and boron abundances are anticorrelated, as would be expected from mixing between the H-burning and outer stellar layers. If, as we suspect, a residue of boron is present in the A-type supergiants, we may exclude a scenario in which mixing occurs continuously between the surface and the deep layers operating the CN-cycle. Further exploitation of the B II 1362 Å line as an indicator of the evolutionary status of A- and B-type stars will require a larger stellar sample to be observed with higher signal-to-noise as attainable with the Hubble Space Telescope.

*Send offprint requests to*: Kim Venn, Univ. Sternwarte-München, Scheinerstr. 1, 81679 München, Germany

## 1. Introduction

The evolution of the massive ($\sim$ 10 to 30 M$_\odot$) stars directly affects the chemical evolution of their host galaxy. These stars are important sites for the nucleosynthesis of the chemical elements, and the distribution of the new material into the interstellar medium. On the theoretical front, the evolution of massive stars is incompletely understood. SN1987A is but one testament to this incompleteness.

In most calculations (e.g., El Eid 1994; Schaller et al. 1992; Stothers & Chin 1991 for M$\geq$10 M$_\odot$; Maeder & Meynet 1989; Chiosi & Summa 1970), a massive star evolves off the main-sequence having experienced some mass loss, but no significant internal mixing. Helium ignition occurs when the star is a blue supergiant, but thermal instabilities cause the star to immediately expand and cool. As a red supergiant, the star has an extensive convective envelope. Then, it returns to the A-type supergiant region for most of its He-core burning lifetime. In this scenario, the convective envelope when the star is a red supergiant will affect the surface chemical abundances, e.g., the carbon, nitrogen, and oxygen abundances are altered at this time by what is commonly called the first dredge-up. An alternative scenario (Stothers & Chin 1991 for M<10 M$_\odot$; Stothers & Chin 1976 and references therein; Chiosi & Summa 1970; Iben 1966) is that He-core burning can be initiated stably while the star is a blue-supergiant. This occurs because of a difference in the H compositional profile; a fully convective intermediate zone produces a H plateau which inhibits the rapidly expanding H-burning shell. Thus, the star does not visit the red supergiant region, and, therefore, does not undergo the first dredge-up. Unless alternative mixing processes occur, stars evolving by this alternative scenario are expected to preserve their surface chemical abundances. Alternative processes have been suggested: (Maeder 1987; Denissenkov 1994; Langer 1994) they include rotationally-induced mixing on or near the main-sequence in massive stars. In these scenarios, partial mixing of the surface with deeper layers can occur, affecting the surface abundances of some elements, but such mixing is activated only when the rotational velocity exceeds a mass-dependent critical velocity.

Abundance analyses of blue supergiants are a useful way to distinguish between the different evolutionary scenarios and provide additional observational constraints for stars near the main-sequence. Venn (1995) has spectroscopically determined abundances for carbon, nitrogen, oxygen, and metals for a se-lection of Galactic A-type supergiants. The results strongly suggest that these stars have *not* undergone the first dredge-up. However, the N/C ratios, and nitrogen abundances do suggest that some *partial* mixing with CNO-cycled layers has occurred in these stars. Analyses of CNO abundances in B-type supergiants by Gies & Lambert (1992) and Lennon (1994) also suggest that some of the evolved stars have been exposed to CN-cycled products. Gies & Lambert have also reported that some non-supergiant B stars show enhanced nitrogen, which suggests that mixing may occur in a few main sequence stars, as predicted by Maeder (1987).

**Table 1.** Synthesized Stars' Parameters and Boron Abundances

| HD | Name | Sp.Ty. | Teff K | Log g | $\xi$ km s$^{-1}$ | $v\sin i$ km s$^{-1}$ | Mass M$_\odot$ | log$\epsilon$(N) ± $\sigma$ NLTE | log$\epsilon$ Si III[a] | log$\epsilon$(B) ± $\sigma$ | log$\epsilon$(B) ± $\sigma$ from BH |
|---|---|---|---|---|---|---|---|---|---|---|---|
| 886 | $\gamma$ Peg | B2 IV | 22670 | 4.0 | 2.6 | 6 | 8.3 | 7.78 ±0.08 | 7.1 | 2.0 ±0.3 | 1.85 ±0.25 |
| 3360 | $\zeta$ Cas | B2 IV | 22180 | 3.9 | 4.9 | 21 | 8.4 | 8.26 ±0.06 | 6.7 | <1.0 | 2.00 ±0.3 |
| 16582 | $\delta$ Cet | B2 IV | 23750 | 4.1 | 4.0[b] | 15 | 8.7 | 8.12 ±0.14 | 6.7 | <2.3 | ... |
| 29248 | $\mu$ Eri | B2 III | 24110 | 3.9 | 4.0[b] | 25 | 9.8 | 7.75 ±0.14 | 6.8 | <2.3 | ... |
| 52089 | $\epsilon$ CMa | B2 II | 20990 | 3.2 | 5.0[b] | 33 | 12.9 | 8.12 ±0.17 | 6.5 | <1.0 | ... |
| 216916 | 16 Lac | B2 IV | 24050 | 3.9 | 4.5 | 13 | 9.8 | 7.66 ±0.11 | 6.9 | <2.0 | ... |
| 46300 | 13 Mon | A0 Ib | 9700 | 2.1 | 4.0 | 17 | 10 | 7.85 ±0.10 | ... | 0.5 ±0.3 | ... |
| 87737 | $\eta$ Leo | A0 Ib | 9700 | 2.0 | 4.0 | 20 | 10 | 8.09 ±0.06 | ... | 0.0 ±0.5 | 2.00 ±0.6 |

[a] Silicon abundances derived from the syntheses in this paper.
[b] From Gies & Lambert, $\xi$=6.1 for HD 16582, $\xi$=9.5 for HD 29248, and $\xi$=15.6 for HD 52089, but we could not fit the profiles with these values.

The light elements – Li, Be, and B – are tracers of exposure to temperatures more moderate than those that process C to N via the H-burning CN-cycle. If even shallow mixing occurs between the surface and the interior, the surface abundances of these light elements must decrease. Mass loss will also accomplish a reduction of the surface abundances. Li and Be are already used in this role in investigations of cool stars (cf, Boesgaard & Lavery 1986; Deliyannis & Pinsonneault 1993), but are unobservable in early-type stars. Boron, which is destroyed when exposed to protons hotter than 4 million K, is observable in hot stars through the ultraviolet resonance line of B II at 1362 Å. Abundances of boron in early-type stars have been determined by Boesgaard & Heacox (1978, hereafter BH) from observations of this resonance line taken with the Copernicus satellite. Spectrum syntheses of 16 stars from 9000 to 25 000 K resulted in mean boron abundance of log$\epsilon$(B)=2.3, which is less than the solar abundance (discussed below). The fact that the boron abundance was apparently uniform across the sample suggests that no significant reduction due to mixing or mass loss has occurred in the observed stars. Boron in early-type stars deserves to be reexamined at this time since significant improvements in stellar atmospheres and line formation calculations have been made, and the atomic data needed for the spectrum synthesis around the B II feature have been improved. Furthermore, significant progress has been made in ultraviolet instrumentation since Copernicus, so that high quality spectra over a large spectral range are now attainable.

In this paper, we discuss the B II 1362.5 Å line in a selection of massive stars to test its effectiveness as an evolutionary indicator. We examine IUE spectra to determine boron abundances in stars of approximately 10 M$_\odot$ from the main-sequence to the A-type supergiant stage, in order to examine the partial mixing scenario for the A-type supergiants and the main-sequence mixing scenario for the B stars.

**Table 2.** SWP Spectra Used in Coaddition

| HD | Sp.Ty. | S/N | SWP |
|---|---|---|---|
| 886 | B2 IV | 26 | 3945,3946,5251,5252,5249,5250,5253. |
| 3360 | B2 IV | 27 | 5408,5409,7806,8133,23862,39620, 3712,3906,5261,5469,5470. |
| 16582 | B2 IV | 26 | 4483,4484,4485,4486,4488,4489,4490, 4491,4492,4493. |
| 22951 | B0.5 V | <10 | 6247,6248,8022,8023. |
| 29248 | B2 III | 15 | 4351,4352,4353,4354,4355,4356,4357, 4358,4359. |
| 31327 | B2 II | <3 | 19632,19633,21279,21280. |
| 34078 | O9.5 V | <5 | 2442,3999,4043,22108,37429. |
| 36959 | B1 V | 8 | 3578,29950,29951,29959,29960,30157, 30158,30159,30206. |
| 41753 | B3 V | <5 | 9933,9934. |
| 46300 | A0 Ib | 25 | 4248,8848,16436,16437. |
| 50707 | B1 IV | 6 | 3575, 27664,27758. |
| 52089 | B2 II | 15 | 35163,35168,36050,36067,35161,35162, 35166,35167,35169,36052,36053. |
| 87737 | A0 Ib | 25 | 3307,4375,4403,16438. |
| 198478 | B3 Ia | <10 | 1823,1824,4503,6335,13907,36937,36938, 38687,38688. |
| 205021 | B2 IV | 11 | 4405,4502,6234,6235,40476,40477. |
| 216916 | B2 IV | 25 | 5354,5360,5361,5355,5356,5357,5358, 5359. |

## 2. Observations

We should like to examine the boron and nitrogen abundances in a selection of B stars in different evolutionary stages. Nitrogen (and N/C) abundances in B stars have been spectroscopically determined by Gies & Lambert (1992) and Lennon (1994), and by Venn (1995) for A-type supergiants. These stars cover a range of masses, but we have chosen our program stars from a limited mass range, $\sim$ 10 M$_\odot$, from these papers for

**Table 3.** Line List for Synthesis of B-Stars

| Elem. | Wavel. Å | $\chi$ eV | Log gf[a] | Comment |
|---|---|---|---|---|
| B II | 1362.46 | 0.00 | 0.013 | op |
| Si III | 1361.60 | 19.02 | 0.17 | op |
| Si III | 1362.36 | 17.72 | −0.55 | op |
| Si III | 1362.95 | 21.88 | −1.55 | op |
| Si III | 1363.46 | 17.72 | −0.20 | op |
| Si III | 1363.50 | 17.72 | −0.68 | op |
| | | | | |
| Ar II | 1362.98 | 16.46 | 0.60 | weaker x2 |
| Mn III | 1361.28 | 5.41 | −1.14 | weaker x2 |
| Fe III | 1363.85 | 6.15 | −3.26 | weaker x3 |
| Zn III | 1362.00 | 18.34 | 0.39 | stronger x3 |
| | | | | |
| Ar II | 1363.61 | 18.06 | −0.80 | |
| V III | 1363.06 | 1.93 | −1.30 | |
| Cr IV | 1363.62 | 19.88 | −0.54 | |
| Mn III | 1361.30 | 8.95 | −1.82 | |
| Fe II | 1362.75 | 2.58 | −0.81 | |
| Fe II | 1363.18 | 2.78 | −2.42 | |
| Fe III | 1362.99 | 6.22 | −3.39 | |
| Fe III | 1363.08 | 14.09 | −2.24 | |
| Fe III | 1363.12 | 8.77 | −3.58 | |
| Fe III | 1363.19 | 8.77 | −2.95 | |
| Fe III | 1363.43 | 6.15 | −3.05 | |
| Fe III | 1363.55 | 6.25 | −3.79 | |
| Fe III | 1363.85 | 14.09 | −1.79 | |
| | | | | |
| Fe II | 1361.42 | 2.58 | −0.50 | fake |
| Fe II | 1361.73 | 2.58 | −0.70 | fake |
| Fe II | 1362.85 | 2.58 | −1.00 | fake |
| Fe III | 1363.30 | 8.77 | −2.95 | fake |
| | | | | |
| Dubious: | | | | |
| Fe III | 1362.55 | 8.77 | −3.24 | |
| Fe III | 1362.55 | 8.76 | −4.42 | |
| Zn III | 1362.52 | 18.10 | 0.60 | |

NOTE – op = Opacity Project values.
[a] Log gf values from Kurucz unless otherwise noted.

**Table 4.** Line List for Synthesis of A-Supergiants

| Elem. | Wavel. Å | $\chi$ eV | Log gf Kurucz | Log gf Used | Comment |
|---|---|---|---|---|---|
| B II | 1362.46 | 0.00 | 0.04 | 0.013 | op |
| Si III | 1361.60 | 19.02 | 0.35 | 0.17 | op |
| Si III | 1362.36 | 17.72 | −0.50 | −0.55 | op |
| Si III | 1362.95 | 21.88 | −1.82 | −1.55 | op |
| Si III | 1363.46 | 17.72 | −0.15 | −0.20 | op |
| Si III | 1363.50 | 17.72 | −0.63 | −0.68 | op |
| Fe II | 1361.37 | 3.27 | −0.32 | −0.67 | op |
| Fe II | 1361.37 | 1.67 | −0.35 | −0.63 | op |
| Fe II | 1361.50 | 2.78 | −2.15 | −1.90 | op |
| Fe II | 1362.27 | 2.83 | −1.42 | −2.40 | op |
| Fe II | 1362.32 | 3.27 | −2.74 | −2.32 | op |
| Fe II | 1362.53 | 2.87 | −2.63 | −4.17 | op |
| Fe II | 1362.75 | 2.58 | −0.81 | −1.96 | op |
| Fe II | 1363.18 | 2.78 | −2.42 | −1.90 | op |
| Fe II | 1363.70 | 2.34 | −2.96 | −6.12 | op |
| | | | | | |
| Fe II | 1362.46 | 4.15 | −2.80 | −3.80 | weaker x10 |
| Fe II | 1361.28 | 0.39 | −3.58 | −2.58 | stronger x10 |
| Fe II | 1361.82 | 2.64 | −3.99 | −2.99 | stronger x10 |
| Fe II | 1361.93 | 2.81 | −3.61 | −2.61 | stronger x10 |
| Fe II | 1363.18 | 2.03 | −3.84 | −2.84 | stronger x10 |
| Fe II | 1361.88 | 2.34 | −3.46 | −2.46 | stronger x100 |
| Fe II | 1362.82 | 2.64 | −4.15 | −1.15 | stronger x1000 |
| Fe II | 1363.82 | 2.54 | −3.74 | −0.74 | stronger x1000 |
| Fe II | 1364.00 | 2.54 | −3.93 | −0.93 | stronger x1000 |
| Fe II | 1363.00 | 2.69 | −3.59 | 0.00 | stronger x1000+ |
| | | | | | |
| Fe II | 1363.10 | 4.15 | −3.88 | | |
| Fe II | 1363.46 | 0.35 | −5.57 | | |
| Fe II | 1363.51 | 2.84 | −5.08 | | |
| Fe II | 1363.54 | 2.34 | −3.04 | | |
| Fe II | 1363.62 | 2.69 | −2.06 | | |
| Fe III | 1362.36 | 6.24 | −3.18 | | |
| Fe III | 1362.55 | 8.76 | −4.42 | | |
| Fe III | 1362.55 | 8.77 | −3.24 | | |
| Fe III | 1362.99 | 6.22 | −3.39 | | |
| Fe III | 1363.99 | 10.46 | −2.57 | | |
| Fe III | 1363.19 | 8.77 | −2.95 | | |
| Fe III | 1363.43 | 6.15 | −3.05 | | |
| Fe III | 1363.55 | 6.25 | −3.79 | | |
| Fe III | 1363.85 | 6.15 | −2.76 | | |
| Ni II | 1362.32 | 6.95 | −3.22 | | |
| Ni II | 1362.42 | 6.95 | −3.42 | | |
| Ni II | 1362.48 | 7.12 | −1.98 | | |
| Ni II | 1361.76 | 6.76 | −1.72 | | |
| Ni II | 1361.89 | 2.87 | −1.50 | | |
| | | | | | |
| Fe II | 1362.56 | 2.58 | ... | −2.20 | fake |
| Fe II | 1362.17 | 2.58 | ... | −2.30 | fake |
| Fe II | 1362.08 | 2.58 | ... | −2.50 | fake |
| Fe II | 1363.38 | 2.58 | ... | 0.00 | fake |

NOTE – op = Opacity Project values.

which nitrogen abundances have already been determined. The B II 1362 Å feature is blended with nearby lines in the A- to B-type star temperature range, so that we selected only sharp lined stars.

High resolution, ultraviolet spectra have been gathered from the IUE archives. For each star, the archives were surveyed for SWP spectra; these spectra were extracted and coadded in order to achieve as high a signal-to-noise as possible. Our preference was for spectra taken with the small aperture since these tended to have a higher signal-to-noise per exposure; however, when few of these were available, and if there were several spectra taken with the large aperture, then we coadded these. Before coadding, the spectra were shifted so that the spectral features lined up as well as possible around 1362 Å; spectra were typically shifted by less than 0.2 pixels.

Of all the IUE spectra examined only 8 stars proved to have a combined signal-to-noise ($\geq 15$) that was deemed adequate to determine the boron abundance or to provide an interesting upper limit. These stars are listed in Table 1. All of the stars that were examined are listed in Table 2, along with the IUE file numbers of the SWP spectra summed.

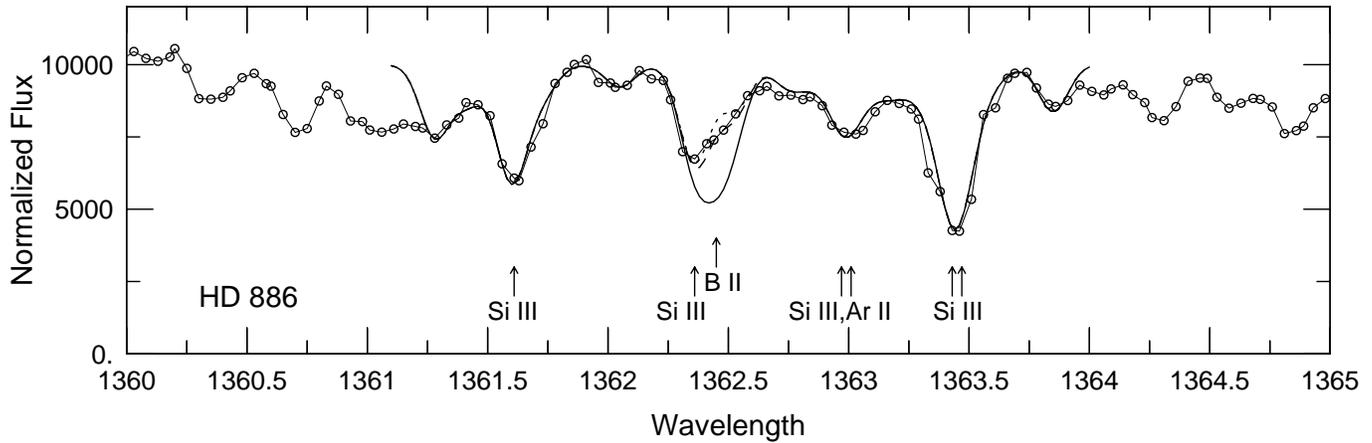

**Fig. 1.** Syntheses for HD 886 with $\log\epsilon(B)=0.0$, 2.0, and 2.9. The synthesis with $\log\epsilon(B)=2.0$ fits the observed spectrum well. The theoretical syntheses are insensitive to boron below $\log\epsilon(B)\sim 1.0$. The dominant features in the spectrum are noted.

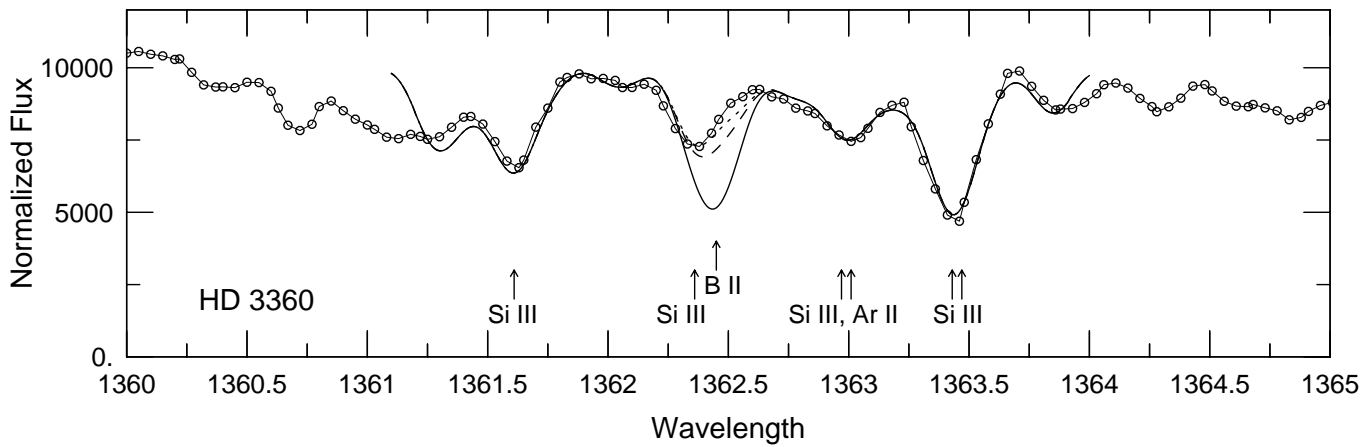

**Fig. 2.** Syntheses for HD 3360 with $\log\epsilon(B)=0.0$, 2.0, and 2.9. The theoretical spectra are insensitive to the boron abundance below $\log\epsilon(B)\sim 1.0$.

### 3. Method

In order to determine the boron abundance in these stars, it was necessary to synthesize the spectrum around the feature because the B II 1362.5 Å line is blended with a strong Si III line (in the B stars) or Fe II lines (in the A-type supergiants, discussed below). The line list provided by BH was augmented in order to synthesize the IUE spectra. Kurucz's (1988) extensive line list was examined carefully for prospective contributors to the 1361 to 1364 Å region. Through repetitive spectrum syntheses, we chose the lines, one by one, that contribute significantly in each temperature class of our target stars. The lines used for the B stars are listed in Table 3. Those used for the A-type supergiants, are listed in Table 4.

Atomic data for these lines were taken initially from the Kurucz line list, which includes wavelength, energy levels, and the gf-values. For the B II line, the energy levels were checked in the Moore tables (1952) and the gf-value taken from the Opacity Project (cf, Cunto et al. 1993). We also did this for the most important features blended with the B II lines, e.g., Si III and Fe II; the wavelengths and energy levels were checked with observed values from the Moore tables, and gf-values taken from the Opacity Project whenever possible.

For a spectrum synthesis, a model atmosphere must be adopted. We chose Kurucz line-blanketed LTE model atmospheres (ATLAS9; Kurucz 1991), with the $T_{\text{eff}}$, log g, and microturbulence ($\xi$) values taken from Gies & Lambert (1992) and Venn (1995). These values are listed in Table 1, along with the masses determined from evolutionary tracks by the authors. For each synthesis, we broadened the theoretical spectrum by 30 km s$^{-1}$ to account for IUE instrumental effects (for IUE's high dispersion camera, R=10 000), and by their projected rotational velocities as listed in Table 1. We then adopted solar Si and Fe abundances, but altered these when necessary to fit the lines *not* blended with the B II feature. In three B stars with high microturbulence values, we could not reproduce the observed spectrum, and found it necessary to reduce the values; in reducing $\xi$, we used the fact that we had several Si III lines of various strengths and found a value that would allow a good fit to these lines. For the A-type supergiants, a solar Fe abundance was determined by Venn (1995), and this value was found to be appropriate for our syntheses.

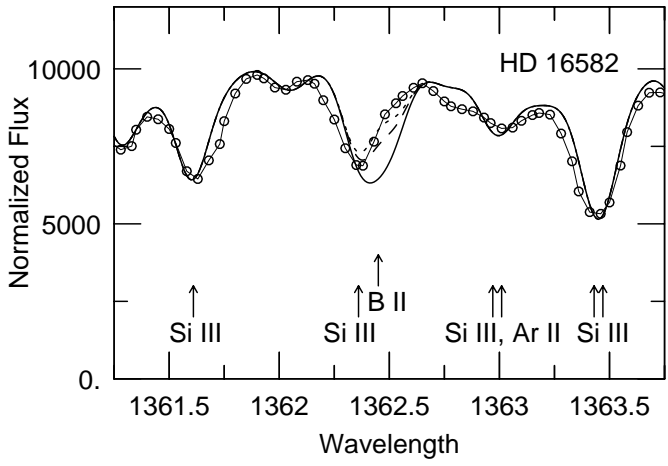

**Fig. 3.** Syntheses for HD 16582 with log ε (B)=0.0, 2.3, and 2.9. The dominant features in the spectrum are noted.

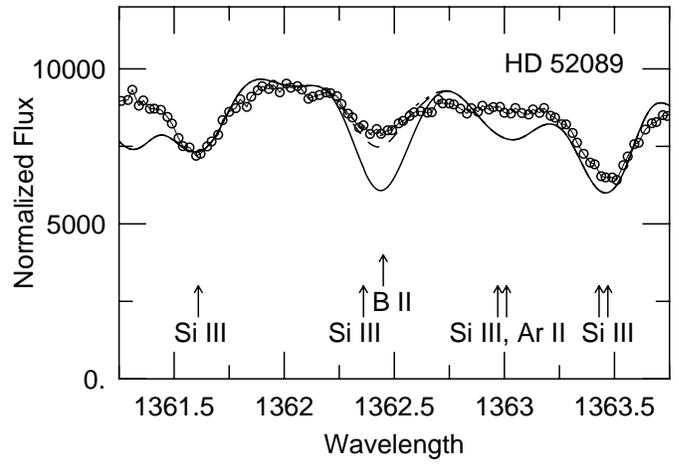

**Fig. 5.** Syntheses for HD 52089 with log ε (B)=0.0, 2.0, and 2.9. The syntheses are insensitive to boron below log ε (B)=1.0.

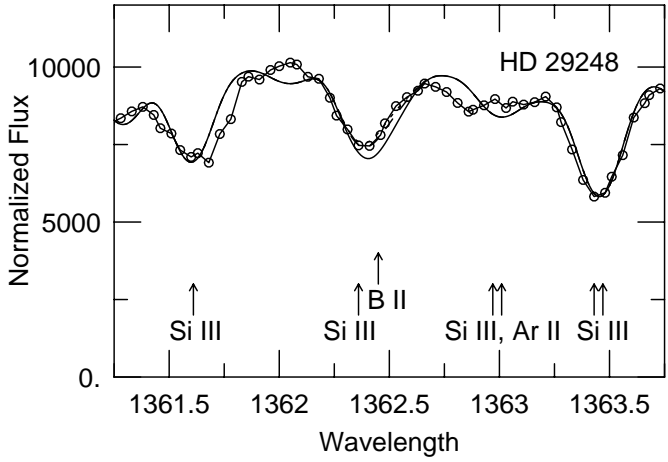

**Fig. 4.** Syntheses for HD 29248 with log ε (B)=0.0, 2.3, and 2.9. We have been forced to reduce the optically determined microturbulence value for this star to fit the observed spectrum. The syntheses are insensitive to boron below log ε (B)=2.3.

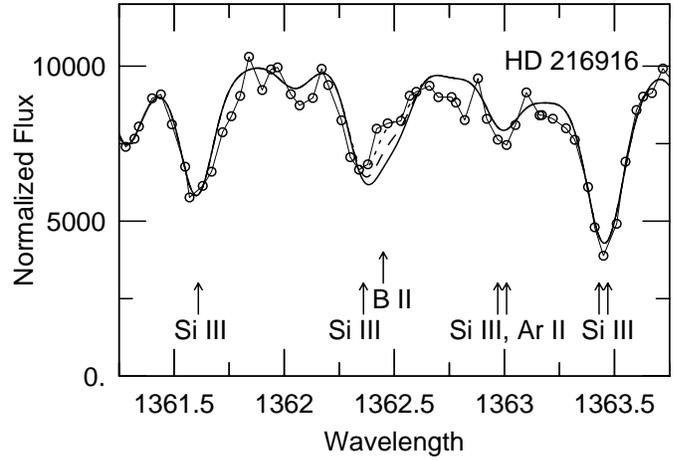

**Fig. 6.** Syntheses for HD 216916 with log ε (B)=0.0, 2.6, and 2.9. The theoretical spectra are insensitive to the boron abundance below log ε (B)∼2.0, which we suggest as an upper limit.

Fortunately, the Si III line that is blended with the B II feature in the B stars belongs to a multiplet contributing several lines near the B II feature. Most of the other Si III lines in the multiplet are clean and unblended, and, hence, define well the Si III contribution to the B II feature. This process ensures that the Si contribution is effectively insensitive to possible errors in the Si III multiplet's gf-value or, equivalently, the Si abundance. We note that the Si abundances we have determined (see Table 1) are approximately 0.7 dex lower than the solar abundances determined by Gies & Lambert (1992) for the optical Si III lines in these stars. This difference in abundance may be due to NLTE effects in the UV lines; optical Si III lines are known to suffer from only very small departures from LTE (Becker & Butler 1990a). Unfortunately, these Si calculations did not include the "prime" series of Si III energy levels (see Becker & Butler 1990b for a description of the model atom), which are needed to explore the NLTE effects of the UV lines (multiplet UV 38, 3d $^3$D − 3d' $^3$P°). No extrapolation of the NLTE calculations to the primed series can be made since the parent term of Si IV for these levels has a higher energy than the ground state for the unprimed series, and could be differently affected by departures from LTE that would subsequently affect the primed Si III levels.

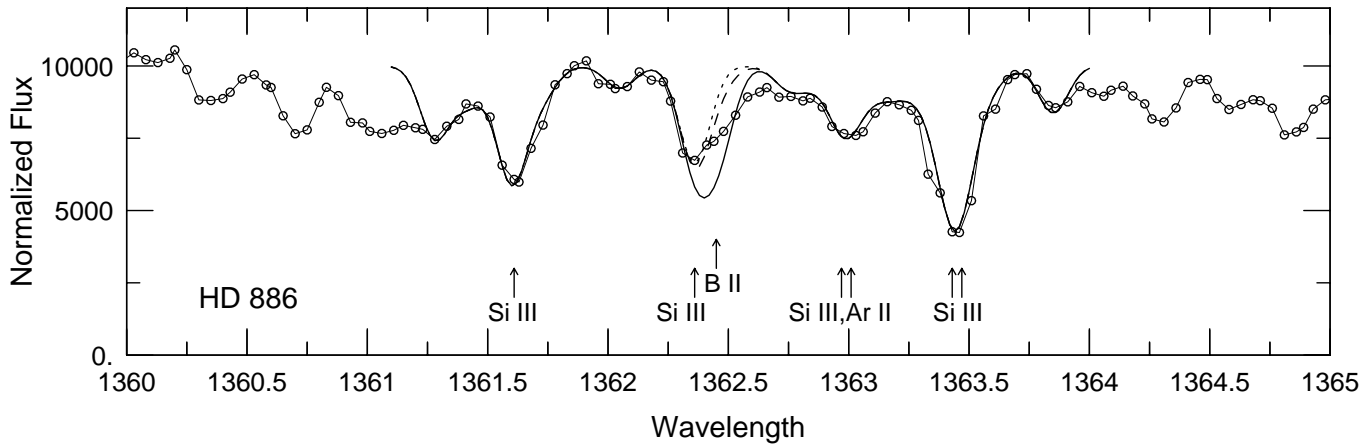

**Fig. 7.** Syntheses for HD 886 without the three dubious (Fe III and Zn III) lines, with log ε(B)=0.0, 2.0, and 2.9. None of these syntheses can reproduce the observed spectrum, thus we surmise that the two Fe III intercombinations lines and the Zn III line must be included.

Unfortunately, but not surprisingly, the Si III line is probably not the only line blended with the B II line. Kurucz lists two Fe III intercombination lines and a Zn III line, see Table 3. The gf-values given by Kurucz cause these lines to be moderately strong. Are these lines real and, if so, are they really so strong? It is difficult to answer this question and confirm Kurucz's atomic data; the Opacity Project considers only LS-coupling for Fe III, therefore there is no information on the two intercombination lines. Ekberg (1993) reported new Fe III laboratory measurements, but not measurements of the intercombination lines. Ekberg's energy levels are accurate to 0.15 cm$^{-1}$, and we find they agree with Kurucz's values to within 0.1 to 0.4 cm$^{-1}$. The resultant change in the wavelengths is less than 0.01 Å. We could not find new information on the Zn III line. If we assume these three lines are accurately represented, then it is possible to fit the spectrum well and determine a boron abundance in the B stars (see Fig.s 1–6). If the Fe III and Zn III lines are omitted from the line list, the fit to the spectra is distinctly inferior − see Fig. 7 showing synthetic and observed spectra of HD 886.

Finally, to make our syntheses of the B stars fit the observed spectra better, we adjusted the Kurucz line list by changing the line strengths of some lines slightly, and adding four "potential" weak lines. This can be done without extreme cause for concern because some of Kurucz's line list was generated from calculations, often involving theoretical, and not observed, energies for some atomic levels, (scaled Hartree-Fock integrals for high configurations of iron group elements; Kurucz 1988), then this can translate to large uncertainties (relative to the widths of the stellar lines) in wavelengths and oscillator strengths. Changes made for the synthesis of the B stars are noted in Table 3, but *no* lines were added that would blend with the B II feature. For the potential lines, the species and other specific atomic data are largely irrelevant as the sample of observed stars spans a narrow range in temperature and surface gravity.

The syntheses of the ultraviolet spectra for the A-type supergiants proved to be more difficult, primarily because the line list from Kurucz does not reproduce the observed spectrum very well (see Fig. 8a). Most Fe II lines included in our synthesis from Kurucz's line list are those from observed energy levels, thus the wavelengths are very accurate, so that we did not adjust any of these. However, new gf-values from the Opacity Project are available for several of the Fe II lines. Changing the gf-values to Opacity Project values whenever possible improved the spectrum synthesis slightly (see Fig. 8b). Given that the Kurucz values could be uncertain by large factors, we then took the liberty of changing some of those gf-values to fit the observed spectrum better (see Fig. 8c); most of these lines were not blended with the B II feature, and thus do not introduce any significant errors into this analysis. However, for a truly excellent fit to an observed spectrum (see Fig.s 8d and 9), we needed to introduce additional or 'phantom' lines into the line list. Since Fe II lines dominate the spectrum, we introduced four phantom Fe II lines and adjusted the gf-values to fit the spectrum; these lines are assumed to be Fe II lines, as yet undetected in laboratory spectra. It is, perhaps, more likely in view of the relatively large required gf-values that these lines may be from elements poorly represented in Kurucz's line list. Three of the four lines are within 0.2 Å of the B II line, but in no way could we adjust the boron line to match the observed spectra before adding these lines.

We have neglected non-LTE effects throughout this analysis. The model atmospheres are built on the LTE assumption. Non-LTE effects in B stars on or near the main sequence are expected to be small (cf, Kudritzki 1988), and have been shown to be small from several line formation calculations (e.g., Becker & Butler 1988a for N II, Becker & Butler 1988b for O II, Eber & Butler 1988 for C II). For the A-type supergiants, Venn's (1995) analysis shows that LTE is an adequate assumption for the model atmospheres and the determination of the iron abundance from Fe II lines. In fact, LTE is expected to be a good assumption in A-type supergiants whenever analyzing the dominant ion of an element, and boron is predominantly in the form of $B^+$ in these stars. Boron in the B-type stars, however, exists primarily as $B^{3+}$ ions with $B^+$ ions as a minor species so that non-LTE effects are possibly noticeable for the B II 1362 Å line.

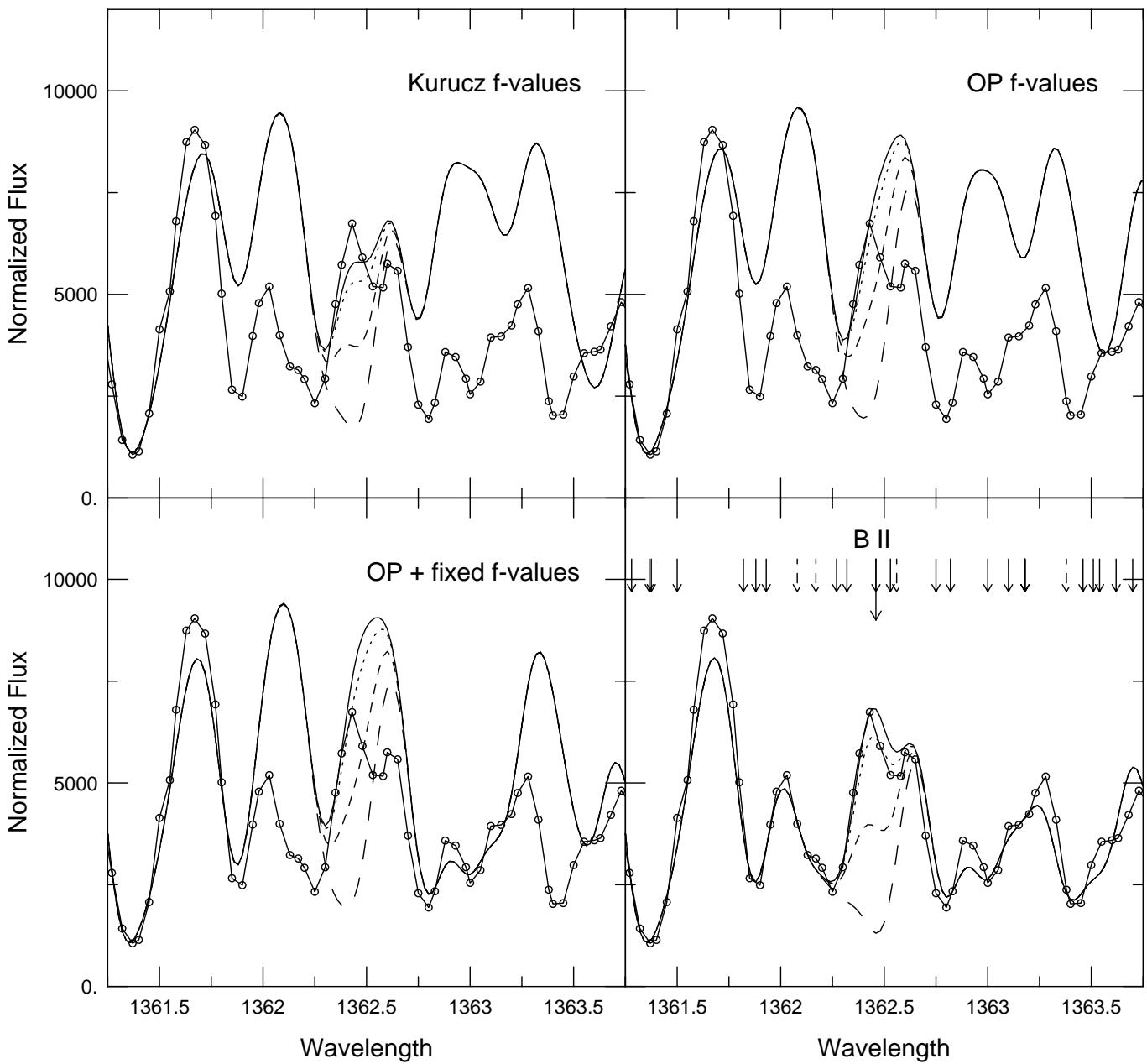

**Fig. 8.** Syntheses for HD 87737 with B/H=0 (solid line), and log$\epsilon$(B)=0.0 (dotted), =1.0 (short dashed), and =2.9 (long dashed). These plots show our increasing success at synthesizing the 1362 Å region; (a) using only Kurucz's lines and gf-values, (b) using Opacity Project gf-values whenever possible, (c) adjusting the line strengths of a *few* of Kurucz's lines, (d) adding the four phantom Fe II lines. All of the Fe II lines in our syntheses are noted by arrows, with dashed arrows representing the four false lines. None of these line lists could be used to reproduce the meteoritic boron abundance. Our syntheses suggest a boron abundance between 0 and 1 boron atom per $10^{12}$ hydrogen atoms.

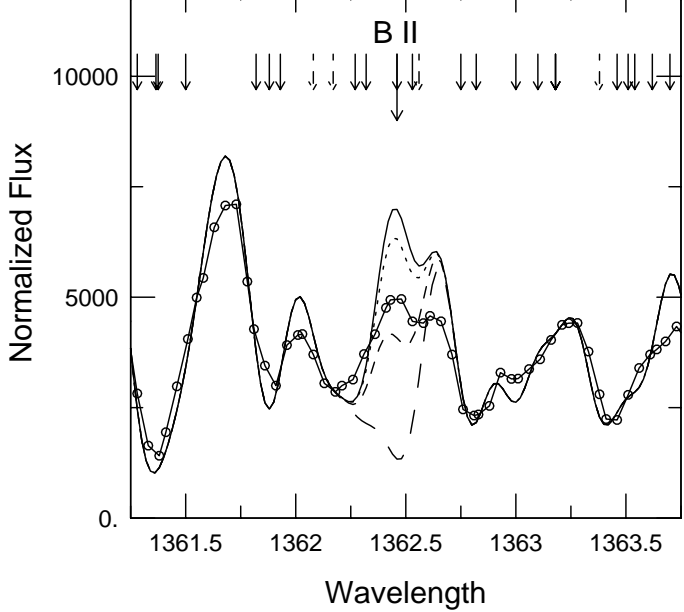

**Fig. 9.** Syntheses for HD 46300 with B/H=0 (solid line), 0.0 (dotted), 1.0 (short dashed), and 2.9 (long dashed). Although there may be another missing feature near 1362.65 Å, our syntheses suggest that between 3 and 10 boron atoms per $10^{12}$ hydrogen atoms are present. The Fe II lines in our syntheses are noted by arrows, with dashed arrows representing the four false lines that we added (see text).

## 4. Boron Abundances

We have determined boron abundances for three stars from IUE spectra, and upper limits to boron in five additional stars. The IUE spectra for the many other B stars that we examined proved to have too low of a signal-to-noise (<15) for a decent synthesis.

Interpretation of the stellar boron abundances depends, in part, on the assumed initial (i.e., undepleted) boron abundance for these stars. With a larger sample and higher quality spectra, it would be possible to infer that value from the stellar abundances. An alternative is to adopt the meteoritic abundance; in their comprehensive review, Anders & Grevesse (1989) give $\log\epsilon(B) = 2.88 \pm 0.04$. A more recent measurement of the meteoritic abundance (Zhai & Shaw 1994) corresponds to $\log\epsilon(B) = 2.78 \pm 0.05$. One may also adopt a solar abundance, although these results are uncertain; an abundance by Kohl et al. (1977) is seemingly consistent with the meteoritic value, although an upper limit by Hall & Engvöld (1975) is lower, based on an excited B I photospheric line. One may also choose a stellar boron abundance as the cosmic value, i.e., the LTE analyses by BH gave $\log\epsilon(B) = 2.3 \pm 0.2$ as the mean of 16 stars across the temperature range 9000 to 25 000 K. Furthermore, Lemke, Lambert, & Edvardsson's (1993) LTE boron abundances, from a B I resonance line in two F stars, are consistent with BH's mean value. Kiselman & Carlsson (1995), however, suggest that correction for non-LTE effects on the B I line will raise the stellar abundance from the F stars, and place it close to the meteoritic value. The independence of the boron abundance on $T_{eff}$ in BH's survey suggests that the non-LTE effects are small, yet these results are a factor of 4 (0.6 dex) less than the meteoritic F-stars. Measurement of the interstellar B II 1362 Å line toward ζ Oph provides some support for the stellar rather than the meteoritic value. When using the reference abundance of $\log\epsilon(B) = 2.3$, Federman et al. (1993) find a depletion factor for the gaseous boron that is similar to that of zinc, an element having a similar condensation temperature. If the meteoritic abundance were assumed, the interstellar boron depletion would be somewhat greater than that of zinc. Adoption of the meteoritic (solar) boron abundance presumes that elemental abundances in objects younger than the Sun ought to be equal to or in excess of the solar abundance. This presumption is not supported by recent analyses of the common elements – C, N, and O – in B stars though; these elements show underabundances relative to solar values of about 0.2 dex (Gies & Lambert 1992; Kilian 1992; Cunha & Lambert 1992, 1994). Although further work is needed to establish the likely initial boron abundance in young stars, the uncertainty is not particularly germane here as we focus on the four stars for which the boron abundance is a factor of 10 (1 dex) or more less than the presumed LTE initial abundance; boron is greatly depleted in these cases even if the initial abundance is taken to be the lower value provided by BH, Lemke et al., or by our analysis of HD 886.

The boron abundances that we have determined are listed in Table 1. Our abundance for HD 886 is in good agreement with that found by BH, and indicates that boron is undepleted or possibly only slightly depleted. Limits on the boron abundance in HD 216916, 29248, and 16582, not analyzed by BH, are also consistent with no or little depletion of boron. By contrast, the abundances for HD 3360, HD 52089, and the A-type supergiants suggest a strong depletion of boron; BH report abundances that are only slightly depleted from solar (−0.6 dex) for two of these stars. Even considering the uncertainties in the IUE spectra (which typically are of a lower quality than BH's Copernicus scans), and the uncertainties in our syntheses, we cannot reproduce the BH abundance for the normal B star, HD 3360. This is because the synthesis causes a deepening of the blend, including the Si III line (see Fig. 2), yet this feature is well constrained by the other Si III lines in our spectrum. We note that BH did not have these other Si III lines because their Copernicus scans covered only a narrow interval around the B II line. As further evidence, we show the IUE spectra of HD 3360 plotted over that of HD 886 in Fig. 10; although, the HD 3360 Si III lines are weaker than in HD 886, the B II feature also appears weaker in HD 3360 suggesting the boron abundance is truly reduced in this star. We admit that the contribution of the B II line to HD 886's spectrum is small for the adopted abundance and that this abundance might be regarded by others as itself an upper limit. A decisive ruling on the presence of boron in these and other B-type stars and their relative boron abundances should be possible from the higher S/N spectra obtainable with the Hubble Space Telescope.

Our negligible boron abundance for HD 87737 is also in stark disagreement with that determined by BH of $\log\epsilon(B)=2.0 \pm 0.6$ – see Fig. 8. BH's brief line list included no Fe II lines which we have found to dominate this region of the ultraviolet spectrum of A0 stars. Omission of the Fe II lines apparently led BH to derive the higher boron abundance. Although we introduced phantom Fe II lines, these do not materially affect the boron abundance that we determine, as can be seen by reviewing Figs. 8a–d. Comparison of the spectra of HD 87737 and HD 46300 shows clearly that the latter has stronger absorption at 1362 Å. Since these stars have identical atmospheric parameters and metallicities to within the errors of measurement, we may conclude that the spectral differences at 1362 Å are due to a higher boron abundance for HD 46300 and cannot be due to a star-to-star changes in the intensities of other – identified or unidentified – lines. Note that the B II line is much stronger for a given boron abundance in these A-type stars than in the B-type stars. This strengthening was noted by BH and is due to the lower temperatures favouring the $B^+$ ion.

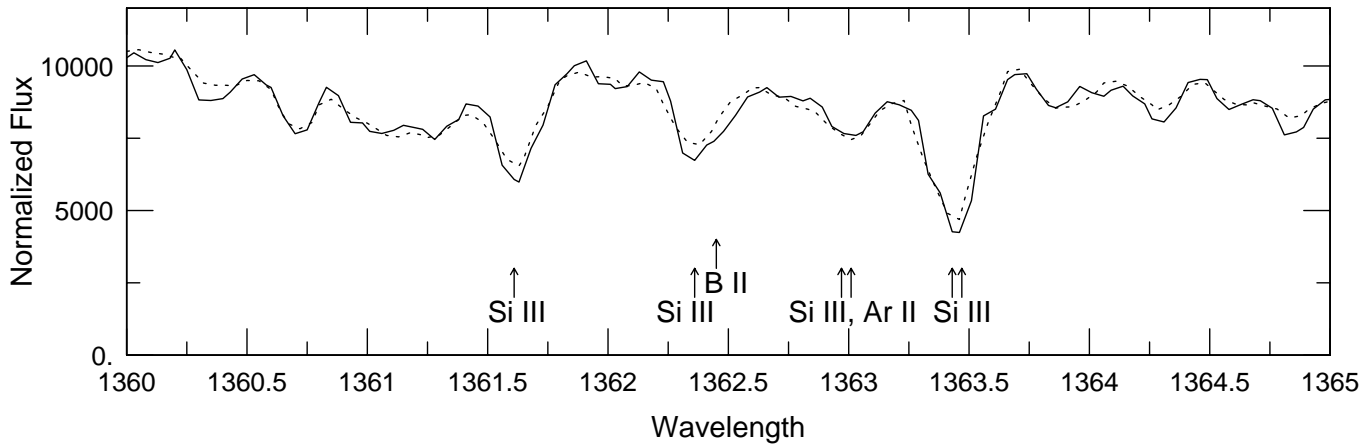

**Fig. 10.** Coadded IUE spectra of HD 3360, compared to that of HD 886, to show that HD 886 has a greater Si abundance and appears to have a somewhat greater boron abundance.

## 5. Boron Abundance Uncertainties

The important parameters to test in our syntheses of the boron line include the model atmosphere parameters ($T_{\rm eff}$, log g, and $\xi$), the Si or Fe abundance, the broadening terms, and the fit of the continuum.

Firstly, we shall discuss uncertainties in the B star syntheses. Test calculations with effective temperatures raised by 500 K, or log gravity raised by 0.2 dex, have a small effect on the syntheses. For HD 886, where we could determine a boron abundance, these changes forced us to raise $\log\epsilon$ (Si) by 0.2 dex and lower $\log\epsilon$ (B) by about 0.3 dex. For the other main-sequence B stars, where only upper limits to the boron abundance were derived, we found the same changes in the Si abundance, but found no significant changes to the boron limits. For the B-giants, these changes in $T_{\rm eff}$ and log g caused no significant changes in the syntheses at all.

The uncertainties in the B star syntheses due to microturbulence are more significant and more difficult to assess. This is primarily because small changes in $\xi$ are countered by small changes in the Si abundance. For example, for HD 886, nearly identical theoretical spectra are derived if $\xi = 4.0$ km s$^{-1}$ and $\log\epsilon$ (Si) $= 6.9$, i.e., $\Delta\xi = +1.4$ km s$^{-1}$ and $\Delta\log\epsilon$ (Si) $= -0.2$ from the values in Table 1. These changes do not affect the B II feature, so that very similar B abundances or upper limits are determined. This is even true for the non-main-sequence B stars, but only after the initial large change in $\xi$ that we were forced to make (discussed above). And, again, we emphasize, that there was no way to synthesize the B-giant spectra without reducing the microturbulent values from the large values determined from the optical lines by Gies & Lambert (1992).

When defining the continuum of the observed B star spectra in order to match them to the theoretical spectra, we have found that an uncertainty of about 3% in the flux can be made depending on the signal-to-noise and which parts of the spectrum are examined. This small difference can then be corrected by a small change in the Si abundance (<0.2 dex) so that those lines fit equally well, but then this can have a significant effect on the boron abundance. For HD 886, lifting the IUE spectrum relative to the theoretical spectrum by 3% forces us to decrease $\log\epsilon$ (Si) by 0.2 dex, but also to reduce $\log\epsilon$ (B) by ~0.3 dex. The upper limits to the boron abundance are similarly affected. Finally, small changes in the rotational or instrumental broadening parameter of less than 3 km s$^{-1}$ have no significant effects on the syntheses of the B stars.

In our examination of the uncertainties in the syntheses of the A-type supergiants, we found that only the microturbulence and $\log\epsilon$ (Fe) had any significant effects. $\Delta\xi = \pm 2$ km s$^{-1}$, or $\Delta\log\epsilon$ (Fe) $= \pm 0.2$ dex, drastically affect all features in the UV spectrum, and one does not compensate exactly for the other. But, even if we permit a very poor fit of the Fe lines, the boron feature does not change a great deal, resulting in an uncertainty of only <0.3 dex in $\log\epsilon$ (B). Changing the fit of the continuum by 3% has a negligible effect on the boron abundance, even though a larger uncertainty in the location of the continuum is possible. In fitting the continuum for the A-type supergiants, we synthesized a larger spectral range, from 1357 to 1367 Å and found a few regions of near continuum (~ 1358 and 1360 Å) for our fit estimates. Finally, changes in $T_{\rm eff}$ of $\pm 300$ K, log g of $\pm 0.2$ dex, or the instrumental and rotational broadening parameters by $\pm 3$ km s$^{-1}$, have no significant effect on the theoretical UV spectrum, and thus on the boron abundances determined (e.g., $\Delta\log\epsilon$ (B) < 0.1 dex).

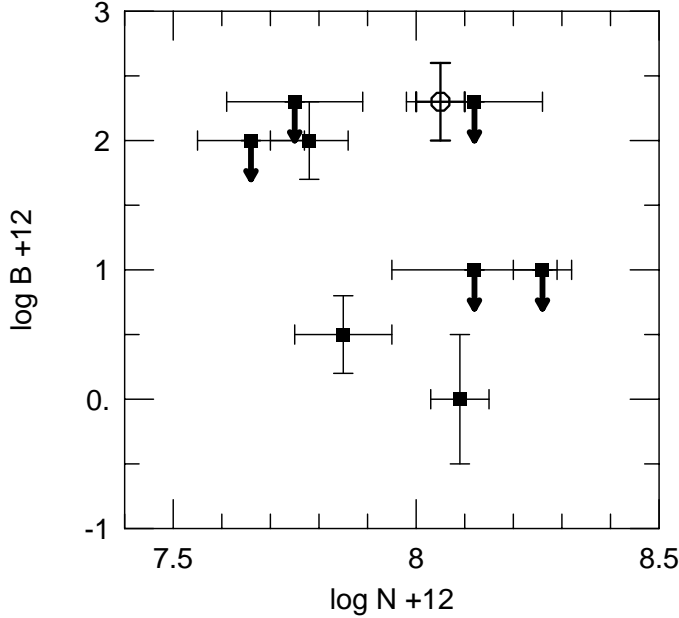

**Fig. 11.** Dubious relationship of boron with non-LTE nitrogen abundances.

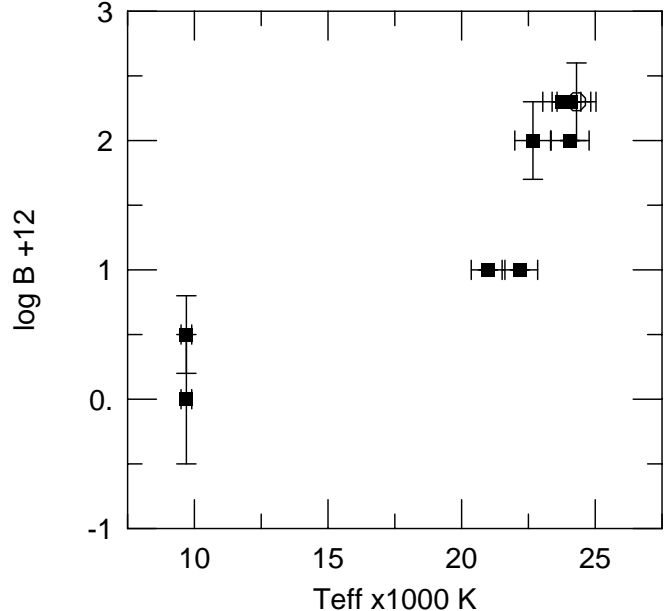

**Fig. 12.** Relationship of boron with temperature, which may indicate some evolutionary effect, but is currently of dubious quality.

## 6. Discussion

This study was undertaken to explore the use of the surface boron abundance as an indicator of the evolution of massive stars. A discussion of our results should be prefaced by the remark that we have found the boron line difficult to synthesize because of the low quality of even the best available IUE spectra, the lack of good atomic data of lines in the 1362 Å region, and the likely presence of unidentified lines that need to be considered in synthesising the spectra of both the B- and A-type stars. If boron is an evolutionary indicator, then we should see a trend in the boron abundance with evolutionary status (i.e., $T_{eff}$ for stars of the same masses) and with the nitrogen, and/or carbon, abundances. There is a hint of a trend in the boron abundances with both the nitrogen abundances and the effective temperatures of these stars, as seen in Fig.s 11 and 12, although with so few stars, and only upper limits on more than half of the boron abundances, the trends are of dubious quality.

It is perhaps of more interest to comment on the boron abundances in selected stars. The lack of boron in HD 3360 is odd since this star is presumably a normal main-sequence B star, similar to the other three, non-giant B stars in this analysis. Comparison of the spectra of HD 886 and HD 3360 (Fig. 10) suggests directly that boron is depleted in the latter star. We also note that the nitrogen abundance in this star is elevated, when compared to either the mean of many B stars or to only HD 886. Perhaps this star has already undergone significant mixing while on the main-sequence. Gies & Lambert (1992) also came to this conclusion for a few main-sequence B stars that show nitrogen enhancements, whereas most of the stars in their sample have He, C, N, and O abundances that are close to Orion Nebula abundances. Maeder (1987) proposed that massive stars ($M \geq 25\ M_\odot$) could undergo rotationally induced mixing *during* their main-sequence lifetime, but that they must rotate above a certain critical velocity. All of the stars chosen for our analysis are sharp lined objects, but this gives no information on their actual rotation rates since they may be pole-on rapid rotators. Thus, HD 3360 might be a rapid rotator that is mixed on the main-sequence, while the other B stars are slower rotators that have not been mixed; or HD 3360 may have already spun down at its surface causing some mixing, while the other B stars are still rapidly rotating; or there may be other unforeseen anomalies for this star. Operation of the CN-cycle to convert C to N occurs at temperatures at which B is rapidly and thoroughly destroyed. Thus, if HD 3360 has a residue of boron at its surface, coexistence of an enhanced nitrogen abundance with a depleted but finite boron abundance sets an obvious constraint on the mixing processes; for example, a steady and complete circulation of material to regions running the CN-cycle is excluded as this would lead to complete loss of boron. Preservation of some boron also limits the integrated mass loss to less than about 2% of the total initial mass. Therefore, a more careful study of boron in this star with better quality spectra is highly desirable.

We also find a very low limit on the boron abundance in the B-giant, HD 52089 (the high limit we found for HD 29248 has no astrophysical significance since it may or may not have undergone any mixing). The fact that boron cannot be near its cosmic value in this star suggests that it may have undergone significant mixing (or mass loss) on or near the main-sequence. However, we cannot rule out non-LTE effects as being the cause of the disappearance of the boron feature in this star, although these effects would have to be a steep function of surface gravity in order to account for the particular weakness of the B II line in this star.

Similarly, the reduced abundance of boron in the A-type supergiants is interesting, especially as the nitrogen abundance determined for these stars is very near solar ($\log \epsilon$ (N)=8.00, Anders & Grevesse 1989), which would seem to suggest that dredge-up as a red supergiant. Although we cannot rule out non-LTE effects on the boron abundance, Venn (1995) remarks that perhaps the solar nitrogen abundance was not the initial value for such young stars. The mean nitrogen abundance in $\sim$ 10 $M_\odot$ B stars (mean of 70 non-supergiant B stars from Gies & Lambert 1992, Kilian 1992, and Cunha & Lambert 1994) is slightly less than solar, by $\sim$0.2 dex. These stars are the natural progenitors of the B- and A-type supergiants. Comparing the nitrogen abundances to the mean of the B stars, then nitrogen is overabundant in HD 87737 and HD 52089 by $\sim$0.3 dex, and only 0.05 dex in HD 46300. The first dredge-up predicts that N increases by a larger amount, $\sim$0.45 dex (c.f., Schaller et al. 1992), which may indicate only partial mixing, but the difference between the prediction and the observational inference may be bridgeable within the errors. Nonetheless, the inferred enrichment of nitrogen indicates that some mixing has occurred with deeper stellar layers, and necessarily this mixing reduces the boron abundance. Furthermore, the mixing appears to have been more severe in HD 87737 than in HD 46300.

Recently, Denissenkov (1994) has shown that a treatment for turbulent diffusive mixing may cause changes in the abundances of CNO in 10 $M_\odot$ stars as they evolve off the main-sequence. The analysis of non-LTE nitrogen and carbon in A-type supergiants (Venn 1995) and low luminosity B-type supergiants (Gies & Lambert 1992, Lennon 1994) support this sort of evolutionary scenario which predicts a low boron abundance.

## 7. Conclusions

Conclusions concerning the role of boron as an indicator of the evolutionary status of A and B type stars must be tempered by the realization that the B II 1362 Å line is not unblended. There is a possibility of unidentified lines contributing to the absorption feature that we attribute to B II. This qualification, however, does not preclude us from drawing several useful conclusions.

A meteoritic/solar boron abundance, $\log\epsilon$(B)=2.8, results in a 1362 B II feature that is too strong to match the observed spectrum of all our program stars. In particular, the predicted feature fails to match the observed one for the two main sequence stars – HD 886 and HD 216916 – that appear unmixed according to their low nitrogen abundances. The maximum boron abundance from our small sample is about 0.8 dex less than the meteoritic/solar value. In this respect, we confirm BH's result that A and B-type stars have a boron abundance, $\log\epsilon$(B)=2.3, less than the solar value. The correct interpretation of this difference is as yet unclear. It may represent a true (natal) reduction, as seems apparent for C, N, and O from the recent analyses of B stars cited above; infall of primordial matter could accomplish these overall reductions. Mass loss by main sequence B stars can also lead to a reduced initial boron abundance if the cumulative loss exceeds about 2 per cent of the initial mass. Unless the extreme outer layers are unexpectedly convective, mass loss will not lead to a steady reduction of the surface boron abundance but rather provide for an abrupt decrease as the layers thoroughly depleted of boron are exposed. This scenario does not fit BH's observation of a uniform boron abundance. Alternatively, the boron abundances from LTE analyses of the spectra of these warm and hot stars may be an underestimate of the true abundance that are now feasible.

Whatever the true initial boron abundance of these stars, and despite the possibility of non-LTE effects, it seems clear that boron is depleted in some stars, and especially in the supergiants. There is a suspicion that the nitrogen and boron abundances are anticorrelated, as would be expected from mixing between the H-burning and outer stellar layers. The magnitude of the nitrogen enrichment (0.3 dex or so) and the boron depletion (a factor of $\sim$100) are roughly consistent with predictions of the first dredge-up. If, as we suspect, a residue of boron is present in the A-type supergiants, we may exclude a scenario in which mixing occurs continuously between the surface and the deep layers operating the N-enriching CN-cycle.

Further exploitation of the B II 1362 Å line as an indicator of the evolutionary status of A and B-type stars will require a larger stellar sample to be observed to the higher S/N ratios attainable with HST. A parallel effort to improve the completeness of the required atomic data should be undertaken. Calculations should also be made to assess the magnitude of the non-LTE effects.

*Acknowledgements.* Special thanks to Margaret Hanson for access to and help with the full IUE archives, and RDAF extraction and reduction programs at the U.Colorado at Boulder. Thanks also to Sylvia Becker for help and advice on Si III and in obtaining and reviewing some Fe III atomic data, and to Danny Lennon for several useful conversations. This research has been supported in part by the National Aeronautics and Space Administration (grant NAG5-2484).